\documentclass[review]{elsarticle}

\usepackage{lineno,hyperref}
\modulolinenumbers[5]

\usepackage{amsmath,amsthm,amssymb,epsfig,bm,amsmath}
 \usepackage{graphicx}
\usepackage{float}
\usepackage{amsmath}
\usepackage{mhchem}
\usepackage{xparse}
\usepackage{adjustbox}
\usepackage{MnSymbol}
\usepackage{mathdots}
\usepackage{makecell}
\usepackage{listings}
\usepackage[printwatermark]{xwatermark}
\usepackage{footnote}
\usepackage[flushleft]{threeparttable}
\usepackage{multirow}
\usepackage{relsize}
\usepackage{lettrine}
\usepackage{courier}
\usepackage{color} 
\definecolor{mygreen}{RGB}{28,172,0} 
\definecolor{mylilas}{RGB}{170,55,241}
\usepackage{amsmath,amsthm,amssymb,amsfonts} 
\usepackage{graphicx} 
\usepackage[margin=1in]{geometry} 
\usepackage[yyyymmdd]{datetime}
\usepackage{algorithm}
\usepackage[noend]{algpseudocode}
\usepackage{lipsum}
\usepackage{setspace}
\usepackage{amssymb}
\usepackage{siunitx}
\usepackage{pdfpages}
\usepackage{booktabs,caption}
\usepackage{braket}
\usepackage[]{hyperref}
\hypersetup{
  colorlinks   = true, 
  urlcolor     = blue, 
  linkcolor    = black, 
  citecolor   = blue 
}
\usepackage[utf8]{inputenc}
 
\usepackage{listings}
\usepackage{color}
\definecolor{codegreen}{rgb}{0,0.6,0}
\definecolor{codegray}{rgb}{0.5,0.5,0.5}
\definecolor{codepurple}{rgb}{0.58,0,0.82}
\definecolor{backcolour}{rgb}{0.95,0.95,0.92}
\lstdefinestyle{mystyle}{
    backgroundcolor=\color{backcolour},   
    commentstyle=\color{codegreen},
    keywordstyle=\color{magenta},
    numberstyle=\tiny\color{codegray},
    stringstyle=\color{codepurple},
    basicstyle=\footnotesize,
    breakatwhitespace=false,         
    breaklines=true,                 
    captionpos=b,                    
    keepspaces=true,                 
    numbers=left,                    
    numbersep=5pt,                  
    showspaces=false,                
    showstringspaces=false,
    showtabs=false,                  
    tabsize=2,
    escapeinside={<@}{@>},
}
 
\usepackage{booktabs} 
\usepackage{siunitx} 

\usepackage{mathtools} 
\usepackage{gensymb} 
\usepackage{mhchem} 
\usepackage{fixltx2e} 
\usepackage{bbm}

\usepackage{multirow}

\usepackage{fancyhdr} 
\theoremstyle{definition}

\theoremstyle{definition}

\theoremstyle{remark}

\usepackage{soul} 
\soulregister\cite7
\soulregister\ref7
\soulregister\pageref7
\usepackage{bm}

\usepackage{framed} 

\usepackage{multicol} 

\usepackage{nomencl} 
\usepackage{tikz}
\makenomenclature
\usepackage[resetlabels,labeled]{multibib}

\renewcommand*\nompreamble{\begin{multicols}{2}}

\renewcommand*\nompostamble{\end{multicols}}

\usepackage{amssymb}
\usepackage{pifont}
\definecolor{light-gray}{gray}{0.95}

\journal{a Jorunal (Review In Progress)}

\begin{document}


\begin{frontmatter}

\title{\large Bayesian Inverse Uncertainty Quantification of the Physical Model Parameters for the Spallation Neutron Source First Target Station}

\author{Majdi I. Radaideh$^*$, Lianshan Lin, Hao Jiang, Sarah Cousineau}

\cortext[mycorrespondingauthor]{Corresponding Author: Majdi I. Radaideh (radaidehmi@ornl.gov)}

\address{Spallation Neutron Source, Oak Ridge National Laboratory, 8600 Spallation Dr, Oak Ridge, TN 37830}

\begin{abstract}
\small

The reliability of the mercury spallation target is mission-critical for the neutron science program of the spallation neutron source at the Oak Ridge National Laboratory. We present an inverse uncertainty quantification (UQ) study using the Bayesian framework for the mercury equation of state model parameters, with the assistance of polynomial chaos expansion surrogate models. By leveraging high-fidelity structural mechanics simulations and real measured strain data, the inverse UQ results reveal a maximum-a-posteriori estimate, mean, and standard deviation of $6.5\times 10^4$ ($6.49\times 10^4 \pm 2.39\times 10^3$) Pa for the tensile cutoff threshold, $12112.1$ ($12111.8 \pm 14.9$) kg/m$^3$ for the mercury density, and $1850.4$ ($1849.7 \pm 5.3$) m/s for the mercury speed of sound. These values do not necessarily represent the nominal mercury physical properties, but the ones that fit the strain data and the solid mechanics model we have used, and can be explained by three reasons. First, the limitations of the computer model or what is known as the ``model-form uncertainty'' that would result from numerical methods and physical approximations. Second, is the biases and errors in the experimental data. Third, is the mercury cavitation damage that also contributes to the change in mercury behavior. Consequently, the equation of state model parameters try to compensate for these effects to improve fitness to the data. The mercury target simulations using the updated parametric values result in an excellent agreement with 88\% average accuracy compared to experimental data, 6\% average increase compared to reference parameters, with some sensors experiencing an increase of more than 25\%. With a more accurate strain response predicted by the calibrated simulations, the component fatigue analysis can utilize the comprehensive strain history data to evaluate the target vessel's lifetime closer to its real limit, saving tremendous target cost and improving the design of future targets as well.

\end{abstract}


\end{frontmatter}


\setstretch{1.5}

\section{Introduction}
\label{sec:intro}

Inverse uncertainty quantification (UQ) is the process of quantifying input parameter uncertainties of a mathematical model using real experimental data with an objective of reducing ad-hoc expert opinion regarding the input parameter physical range \cite{smith2013uncertainty}. The inverse UQ is usually performed using the Bayesian framework due to its robustness in parameter inference as demonstrated by the statistics \cite{castillo2015bayesian} and machine learning \cite{mackay1995bayesian} areas. The value of inverse UQ can already be found in many disciplines, including nuclear engineering \cite{wu2018inverse, domitr2021use} to calibrate nuclear thermal-hydraulics codes, in aerospace \cite{reed2018model} to characterize complex damage scenarios of aerospace components, in nuclear fuel materials \cite{che2021application} using variational Bayesian, in composite science \cite{balokas2021data} to quantify uncertainties of a carbon fiber reinforced composite, in multiphase computational fluid dynamics \cite{liu2021uncertainty}, and many other examples. For the efforts focusing on the theory, the seminal work by \cite{kennedy2001bayesian} introduced the Bayesian inverse framework to calibrate computer models, which has been used as the major building block for the next inverse UQ efforts. The work by \cite{faes2019multivariate} used a multivariate interval approach to conduct inverse UQ with limited experimental data. The work by \cite{yang2022bayesian} introduced a new entropy-based metric by utilizing the Jensen–Shannon divergence to replace the existing distance-based metrics. The entropy-based metric seems to be efficient for mixed uncertainty problems of aleatory and epistemic uncertainty. Additional efforts describing the use of non-intrusive surrogate models to facilitate uncertainty analysis are worth to be highlighted such as deep neural network and support vector machines \cite{radaideh2019combining}, Gaussian processes in the context of Bayesian inversion \cite{wu2018inverse, yu2021efficient, duan2022non}, and deep Gaussian processes  \cite{radaideh2020surrogate}. Examples of frameworks supporting Bayesian modeling and inversion are UQLAB in MATLAB \cite{marelli2014uqlab} and PyMC3 in Python \cite{salvatier2016probabilistic}.

The Spallation Neutron Source (SNS) at Oak Ridge National Laboratory is currently the most powerful accelerator-driven neutron source in the world \cite{mason2006spallation}. The intense proton pulses strike on SNS’s mercury target to provide bright neutron beams, which also lead to severe fluid-structure interactions inside the target \cite{futakawa2000pressure}. Prediction of resultant loading on the target is difficult, particularly when the helium gas is intentionally injected into mercury to reduce the loading and mitigate the pitting damage on the target’s internal walls \cite{liu2018strain, bloklandmeasurements}. The availability of the first target station at the SNS is mission-critical to provide a world-class neutron science program. However, premature failure of target modules has led to several interruptions to the SNS user program and caused a delay in meeting design power, highlighting the need for robust target design. This highlights the first motivation of this work, which will leverage measured strain data for the target in extreme environments to predict and extend the lifetime of future target designs through improving simulation accuracy. The value of modeling and simulation of the mercury target has already been demonstrated by \cite{riemer2005benchmarking} using ABAQUS. 

The second motivation of this study is introducing Bayesian inversion to accelerator physics areas, especially the one focusing on the spallation target analysis. We noticed a scarcity of advanced computational studies in this area. Based upon previous efforts, classical one-at-a-time adjustments of the input parameters have been performed by \cite{lin2021sensitivity,lin2019tunable}, which according to the authors, can be tedious once more experiments become available and the model has more uncertain parameters to tune. These methods also provide only a single optimum point that does not account for the inherent parametric uncertainty. Therefore, in this work, we conduct an inverse UQ analysis using the Bayesian framework to quantify the uncertainty in three major simulation parameters: the tensile cutoff threshold, the density of mercury, and the mercury speed of sound. First, we develop high-fidelity structural mechanics simulation of the mercury target using the Sierra/Solid Mechanics code. Second, we collect real measured strain data resulted from proton pulses on the target. The inverse UQ methodology will quantify the prescribed input parameter uncertainties that minimize the discrepancy between the simulation and the measured data. Given the intensive computational cost of the target simulation, we develop an accurate and fast surrogate model based on polynomial chaos expansions. In addition, given the complexity of the posterior distribution of the input parameters, we apply an advanced Markov Chain Monte Carlo algorithm to generate samples from the posterior distribution. All efforts in this work will yield: (1) a computational UQ and calibration methodology for the accelerator physics area, (2) a better understanding of the input parameter uncertainties related to the mercury target simulation, and (3) an improved target simulation accuracy compared to experimental data.  

For the remaining sections of this work: In Section \ref{sec:sns}, the spallation neutron source facility is introduced. Then, the simulation and experimental setups of the mercury target are described. In Section \ref{sec:method}, the inverse UQ methodology is formulated, which includes the surrogate modeling via polynomial chaos expansions, Bayesian inversion, and Markov Chain Monte Carlo sampling. The inverse UQ results are presented and discussed in Section \ref{sec:res}, while the conclusions of this work are highlighted in Section \ref{sec:conc}. 

\section{Simulation and Experimental Setup}
\label{sec:sns}

\begin{figure}[!h]
    \centering
    \includegraphics[width=0.8\textwidth]{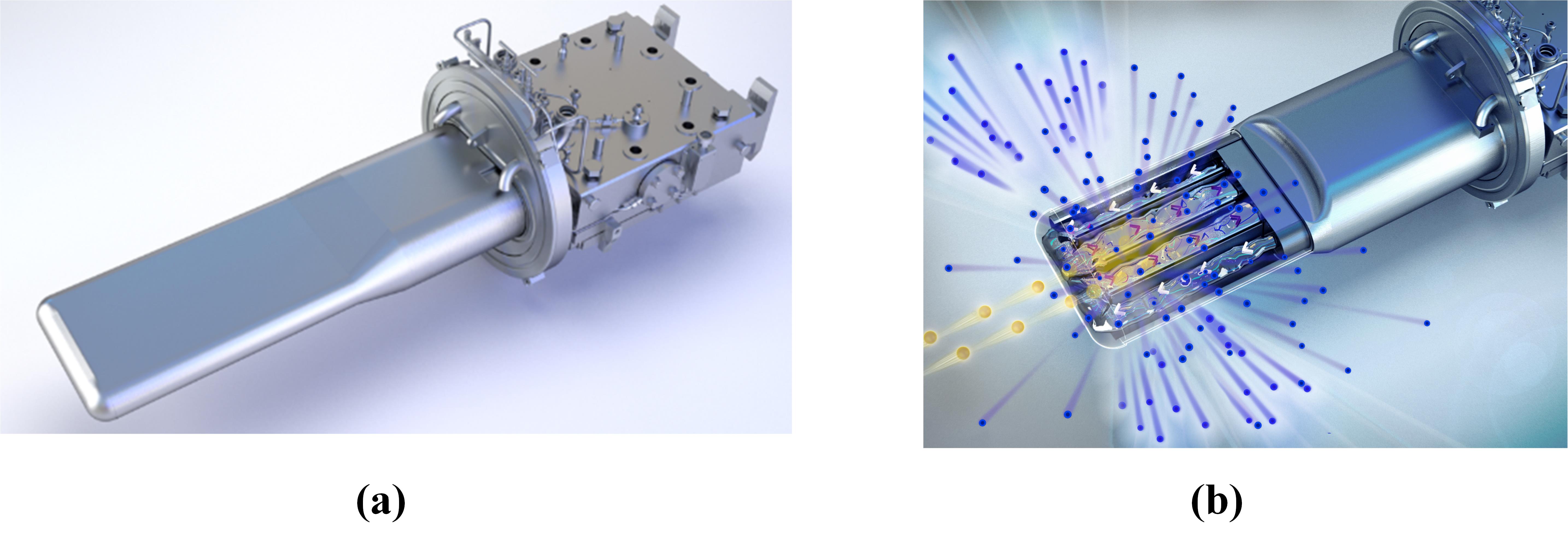}
    \caption{(a) Geometrical sketch of the Mercury target \cite{winder2021evolution}, (b) spallation reaction of proton beam hitting the flowing Mercury \cite{rumsey2018}}
    \label{fig:target}
\end{figure}

The SNS is a very complex facility, consisting of many pieces of equipment, each vital to the operation. This work focuses on the SNS target, which is shown in Figure \ref{fig:target}. The SNS target system operates at high power ($>$1 MW) and uses short pulses (0.7 $\mu$s) of high energy protons ($>$1 GeV). The protons generate neutrons through spallation reactions in a target material, where the generated neutron beam is then used to carry scientific research. The SNS uses liquid mercury as the target material, which has multiple advantages: (1) extended target lifetime, (2) a compact design, and (3) easier cooling as the liquid mercury can be pumped and cooled without a secondary cooling system \cite{winder2021evolution} compared to other solid targets such as Tungsten \cite{rakhman2018proton}.

To reveal the internal structure and the mercury flow, Figure \ref{fig:target}(b) shows a cut view. The mercury flows from the vessel sides and makes a U turn, where the proton beam (the yellow particles) strike the target. As a result of the reaction, the neutrons (blue particles) are generated and the flowing mercury takes away most of the heat generated by the proton beam, is cooled by flowing through a heat exchanger, and pumped back to the target in a continuous cycle.

Nevertheless, the usage of liquid mercury brings potential challenges, including removing the energy deposited by the beam, material compatibility with the liquid metal, and beam-induced thermal shock \cite{riemer2005benchmarking}. The thermal shock is caused by the short duration and large intensity of the proton beam, causing an instantaneous heating that generates pressure in the mercury. The pressure field travels through the fluid and surrounding structure. The damage to structures from cavitation of the mercury was identified as a potential challenge to reliability and the lifetime of the spallation target during the facility design stage. Not surprisingly, target failures due to cavitation damage adversely impacted the facility’s ability to ramp up the beam power to the design level of 1.4 MW for a period of few years, and caused significant unplanned down time to the facility, impacting the user program. Recently, cavitation damage has been greatly reduced with the introduction of helium gas bubble injection into the mercury. 

\subsection{Target Simulation}
\label{sec:sim}

\begin{figure}[!h]
    \centering
    \includegraphics[width=0.9\textwidth]{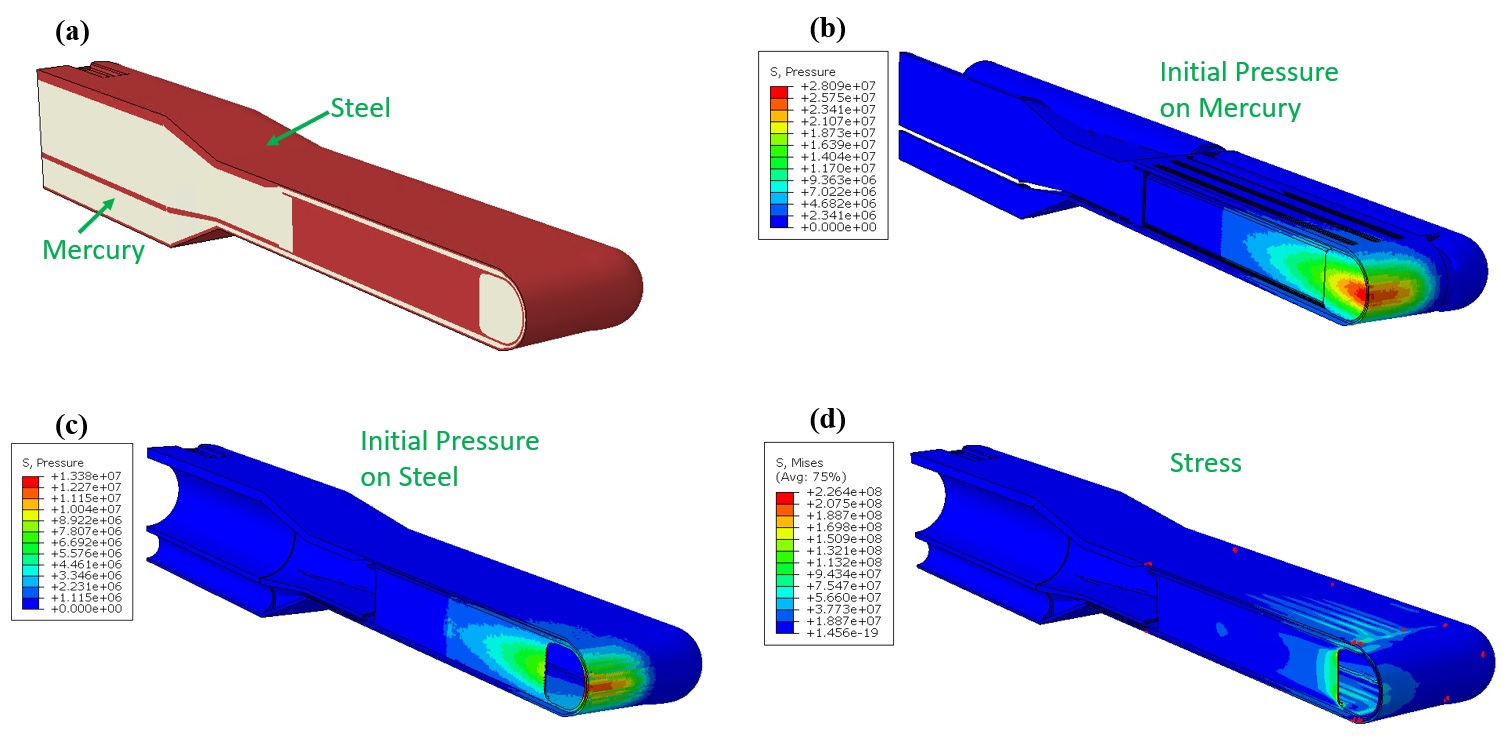}
    \caption{Finite element simulation of the pulsed mercury target: (a) target geometry, (b) initial pressure field on mercury (c) initial pressure field on steel, and (d) stress field.  }
    \label{fig:bayes_target}
\end{figure}

A half symmetric finite element model has been created to simulate the dynamic response of the mercury target due to the proton pulse loads, see Figure \ref{fig:bayes_target}(a). ABAQUS had been used originally to model the target, but in this work, Sierra/Solid Mechanics (Sierra for short) \cite{team2011sierra} was used for modeling and simulation. As Sierra is a government-owned non-commercial software, restrictions on the number of parallel licenses do not exist. 

The proton energy of each pulse deposits on both the mercury and steel parts, being converted into the initial pressure that drives the propagation of stress waves internally. Figure \ref{fig:bayes_target}(b) illustrates the contour of the initial pressure on mercury, while Figure \ref{fig:bayes_target}(c) illustrates the contour of the initial pressure on steel. Obviously, the front part of the target experiences the most extensive initial proton pulse pressure than other target locations. In the target’s finite element model, an equation of state EOS material model is adopted for mercury, with a tensile cutoff value included to represent its cavitation behaviour \cite{riemer2005benchmarking}. Figure \ref{fig:bayes_target}(d) demonstrates the stress response at the specific moment after the proton pulse initiates. The red dots in Figure \ref{fig:bayes_target}(d) represent strain sensors, where the strain responses are recorded. We did not model exact strain sensor, but we did simulate the strain responses at the sensor locations. Some of these sensors are used in this work to record measured data and will be described in the next subsection. 

Current mercury EOS material model uses the key parameters set: mercury density, sound speed, and tensile cutoff threshold. A reference set was reported by \cite{riemer2005benchmarking} to produce a best overall agreement between experimental measurement and finite element simulation for tests without helium gas injection. The tensile cutoff threshold (in Pa) represents the cavitation behavior of mercury bubble excited by the high energy proton pulse, which lost its volumetric stiffness when the mercury bubble grows large enough. The density (in kg/m$^3$) and sound speed (in m/s) values are nominal ones for standard mercury heavy metal, that could vary a little bit when the mercury cavitation appears. In this case we need Bayesian inverse UQ to quantify the parametric values and their uncertainty range for the standard EOS mercury model in order to: improve the current parameter reference set by improving simulation accuracy, and explore the possibility to improve the current EOS model for new target designs.

\subsection{Experimental Data}
\label{sec:data}

\begin{figure}[!h]
    \centering
    \includegraphics[width=0.8\textwidth]{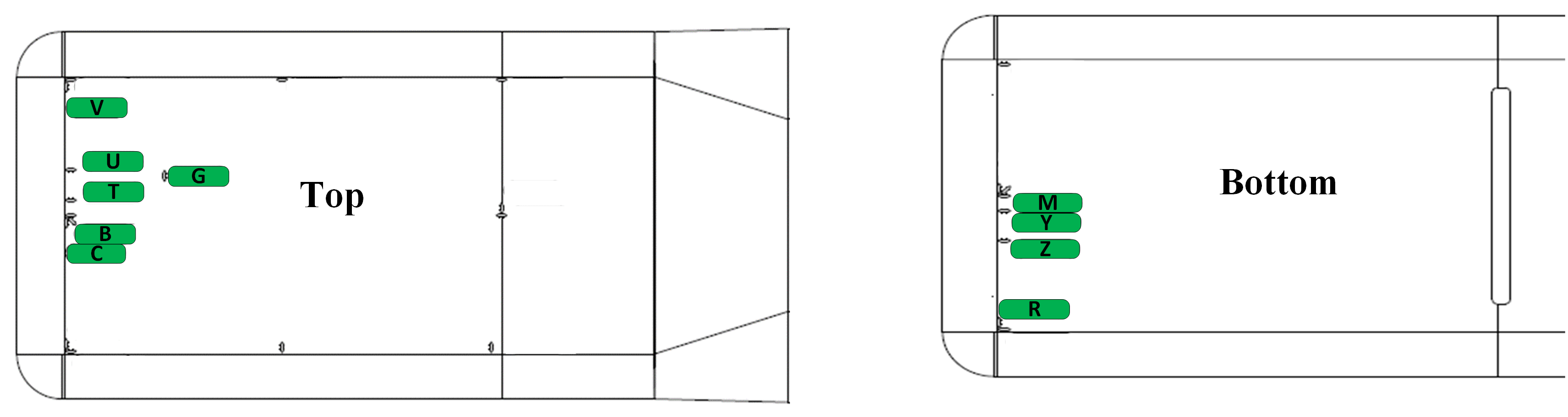}
    \caption{Sensor locations in the mercury target top (left) and bottom (right)}
    \label{fig:bayes_sensors}
\end{figure}

As already described, during the pulsed experiment, an intense proton beam pulse strikes the mercury target from the front (see Figure \ref{fig:target}), and deposits energy into the mercury and the steel vessel. On the external surface of the steel vessel, several strain gauges are attached to collect the strain response. Measurements from these sensors provide an indication of the operational status and fatigue life of the target, as well as benchmarking data for modeling and simulation, which is how we utilize them in this work. Figure \ref{fig:bayes_sensors} shows 10 selected sensors for this study, distributed on the target top and bottom portions. The reader can notice that we focused on the sensors close to the target front where the beam hits, since these sensors observe higher quality signals than the sensors on the back whose signals dominated by noise. Our selection of the sensors were based on previous observations and expert opinions \cite{riemer2005benchmarking}. 

\begin{figure}[!h]
    \centering
    \includegraphics[width=0.65\textwidth]{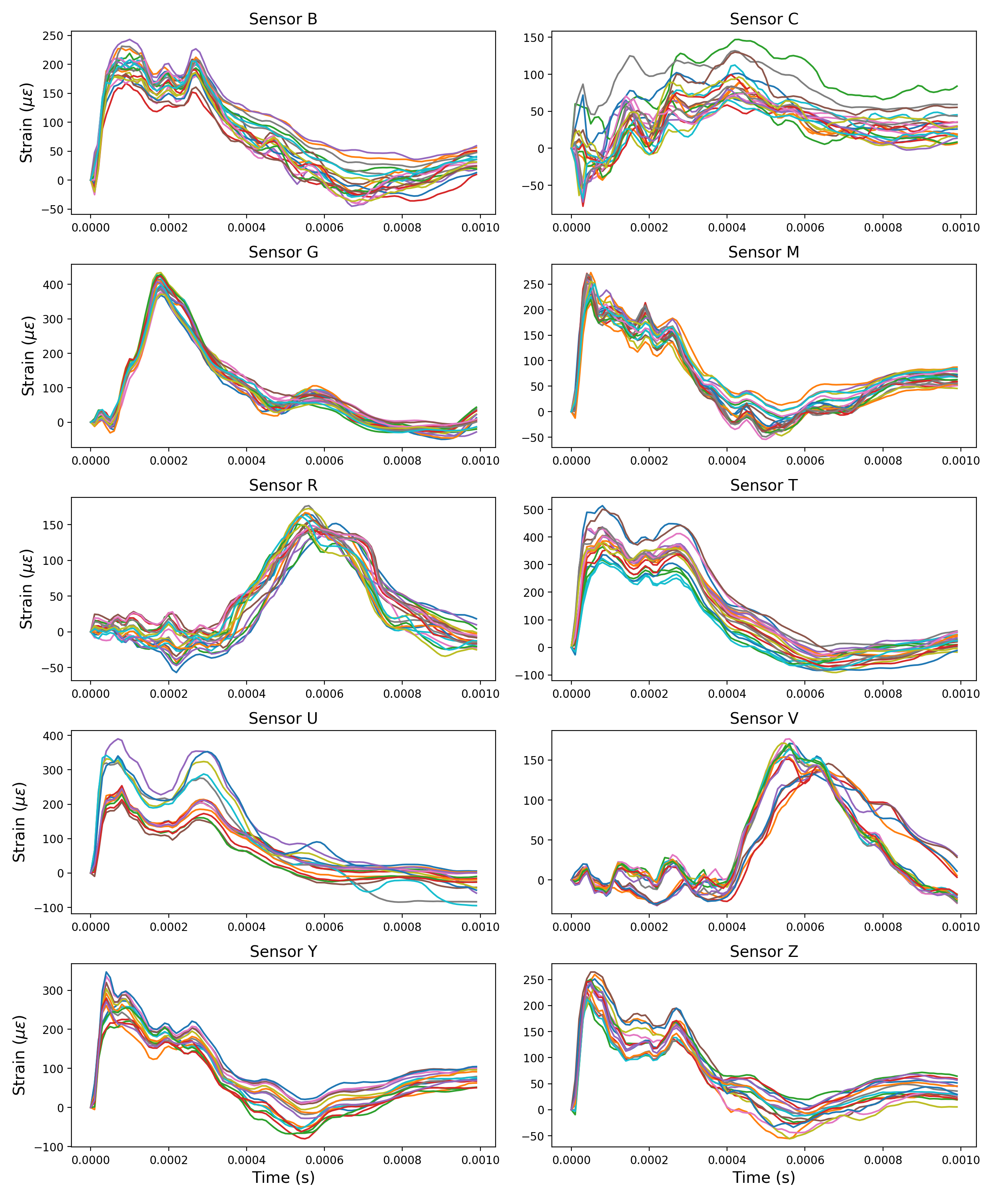}
    \caption{Plots of strain curves for 10 sensors measured at 15-20 different samples/readings of 1 ms long. Each curve refers to a different sample. Sensors (B, C, G, M, R, T) have 20 samples available, while sensors (U, V, Y, Z) have only 15 samples each.}
    \label{fig:bayes_data}
\end{figure}

Figure \ref{fig:bayes_data} illustrates the strain curves as measured by the 10 sensors shown in Figure \ref{fig:bayes_sensors}. The strain unit is micro strain ($1$ $\mu \varepsilon$ = $\varepsilon * 10^6$). The data were collected from four different mercury target designs labelled as: T24, T26, T27, T28, with a beam power level of 1.4 MW for 5-10 consecutive pulses. We can notice that the strain trend differs depending on the sensor location. For example, sensors B and T, which are close to the center where the proton pulse hits, have their strain peak early in time. However, the sensors far from the center such as sensors V and R have a delayed strain peak (see Figure \ref{fig:bayes_sensors}). In addition, the sensors vary in their measurement accuracy as can be observed from the spread of the curves for each sensor. For example, we can notice very good repeatability for sensors G and M, while a fair spread can be seen for sensors U and C. Nevertheless, we should highlight that the strain trend is consistent for each sensor, since we removed all faulty pulses from the sample set that exhibit a noisy signal. These faulty signals were removed as they look like a white noise and extremely far from the rest in Figure \ref{fig:bayes_data}. 

The data structure can be described as follows. Each strain curve has 100 time steps. For sensors \{B, C, G, M, R, T\}, there are 20 samples/curves available, while for sensors \{U, V, Y, Z\}, there are 15 samples/curves. To expand the sample size, we have also calculated the mean curve and the lower/upper bounds (95\% confidence interval) and added them to each sensor. This makes the total number of experimental data points as 
\begin{equation}
    N_D = \underbrace{100*20*6}_{(B, C, G, M, R, T)} + \underbrace{100*15*4}_{(U, V, Y, Z)} + \underbrace{100*3*10}_{\text{(Mean, 95\% Lower, 95\% Upper)}} = 21000.
\end{equation}
This is a decent number of data to perform inverse UQ for the inference of 3 simulation parameters: the tensile cutoff threshold, the density of mercury, and the mercury speed of sound. Indeed, following our analysis and results in section \ref{sec:res}, we noticed that adding the mean and lower/upper bounds to the dataset did not change the conclusion or the results.  

\section{Bayesian Inverse UQ}
\label{sec:method}

In this section, the inverse UQ methodology is described including surrogate modeling, Bayesian inversion, sampling from the posterior distribution, and the computational tools we have used. First, we should define the problem variables before moving on:
\begin{itemize}
    \item The model inputs ($\bm{x}$) are the tensile cutoff threshold, the density of mercury, and the mercury speed of sound. Therefore, the total number of uncertain model inputs is $n_x=3$. 
    \item The computational model is Sierra and it is referred to by $\mathcal{F}$.
    \item The model outputs ($\bm{y}$) are the strain values at 10 sensor locations. Each sensor data is a time series of 100 time steps. Therefore, the total number of model outputs is $n_y = 1000$. 
\end{itemize}

\subsection{Polynomial Chaos Expansions (PCE)}
\label{sec:pce}

Consider a problem with $n_x$ uncertain inputs, and we generated random samples of these inputs $\bm{x} \in \mathbb{R}^{n_x}$ from a joint probability density function $f_{\bm{x}}$. In this context $n_x=3$. Now, for a computer model $\mathcal{F}$, the expectation of the model outputs $\bm{y}$ can be written as 
\begin{equation}
    E[\bm{y}^2]=\int_{R_{\bm{x}}} \mathcal{F}(\bm{z})^2 f_{\bm{x}} (\bm{z})d\bm{z}, 
\end{equation}
where $R_{\bm{x}}$ is the support/domain of the uncertain inputs $\bm{x}$. To create a surrogate for the computer model $\mathcal{F}(\bm{x})$ using PCE, we can write 

\begin{equation}
\label{eq:pce}
    \bm{y}=\mathcal{F}(\bm{x}) \approx \mathcal{F}^*(\bm{x}) = \sum_{\bm{\alpha} \in \mathcal{A}} y_{\bm{\alpha}} \Psi_{\bm{\alpha}}(\bm{x}),
\end{equation}
where $\Psi_{\bm{\alpha}}(\bm{x})$ are multivariate polynomials orthonormal with the respect to the probability density function $f_{\bm{x}}$, $\bm{\alpha} \in \mathcal{A}$ is a subset of selected multi-indices of the multivariate polynomials, and $y_{\bm{\alpha}}$ are their corresponding polynomial coefficients. The PCE surrogate is represented by $\mathcal{F}^*(\bm{x})$. The reader can notice that for practical applications, we truncate the sum in Eq.\eqref{eq:pce} to a finite sum by using only a subset of the multi-indices $\bm{\alpha} \in \mathcal{A}$, which belong to the full index space $\mathcal{A} \in \mathbb{N}^{n_x}$. This explains the approximate sign ($\approx$) in Eq.\eqref{eq:pce}. The multivariate polynomials $\Psi_{\bm{\alpha}}(\bm{x})$ can then be assembled as the tensor product of their univariate polynomials ($\phi$)
\begin{equation}
    \Psi_{\bm{\alpha}}(\bm{x}) = \prod_{i=1}^{n_x} \phi_{\alpha_i}^{(i)}(x_i),
\end{equation}
where the orthonormality of the multivariate polynomials has been verified before \cite{blatman2010adaptive}. Consequently, we can write
\begin{equation}
    \Big < \Psi_{\bm{\alpha}}(\bm{x}), \Psi_{\bm{\beta}}(\bm{x})\Big > = \delta_{\bm{\alpha\beta}},
\end{equation}
where the orthonormal relationship between two univariate polynomial basis can be written as

\begin{equation}
    \Big < \phi_j^{(i)}(x_i), \phi_k^{(i)}(x_i) \Big > =  \int_{R_{\bm{x}}}  \phi_j^{(i)}(x_i) \phi_k^{(i)}(x_i)f_{x_i}dx_i=\delta_{jk}, 
\end{equation}
where $\delta_{jk}$ is the Kronecker delta function and  $\delta_{\bm{\alpha\beta}}$ is its extension to multi-dimensional space. 

The next step is how to determine the values of the polynomial coefficients to build the PCE surrogate. There are many methods to solve this problem such as projection methods (e.g. Gaussian quadrature), ordinary least-square methods, subspace pursuit, Bayesian compression, or least angle regression, where the latter is used in this work. The least angle regression modifies the classical least square minimization by adding a penalty term to facilitate sparsity in high-dimensional problems. Also, least angle regression offers a fast computing time compared to other methods. For conciseness, the details of the least angle regression method can be found in \cite{blatman2011adaptive}. The metric used during the training of the PCE model is the leave-one-out (LOO) cross-validation error ($\epsilon_{LOO}$). This metric limits overfitting of the PCE model to the training data by building multiple PCE metamodels, each with a reduced training set, and comparing its prediction on the excluded point(s). This could help the PCE model to generalize better to new data. The LOO cross-validation error can be expressed as \cite{blatman2010adaptive}
\begin{equation}
    \epsilon_{LOO} = \sum_{i=1}^{N_{val}}\Bigg(\frac{ \bm{y}^i-\bm{y}^{i}_{PCE}}{1-h_i}\Bigg)^2\Bigg/\sum_{i=1}^{N_{val}} (\bm{y}^i-\bm{\bar{y}})^2
\end{equation}
where $N_{val}$ is the number of samples in the validation set, $\bm{y}^i$ is the target output value of the real computer code (e.g. Sierra) of the validation sample $i$, $\bm{y}_{PCE}^{i}$ is the PCE prediction of the validation sample $i$, and $\bm{\bar{y}}$ is the sample mean of the validation set responses. The value of $h_i$ is the $i^{th}$ entry of the vector $\bm{h}$
\begin{equation}
    \bm{h} = diag(\bm{A}(\bm{A}^T\bm{A})^{-1}\bm{A}^T),
\end{equation}
where the regression matrix $\bm{A}$ contains the evaluation of the
basis polynomials at every validation set point

\begin{equation}
    A_{ij} = \Psi_j(\bm{x}^{(i)}), \quad i=1, .. , N_{val}, \quad j= 0 ,.. , p-1,
\end{equation}
where $p$ is the truncated polynomial order. 

\subsection{Bayesian Statistics}
\label{sec:delta}

Bayesian statistics is a popular framework for inference that allows us to fit a statistical model by combining the prior knowledge of the uncertain parameters with new observed data (e.g. from experiments) using the Bayes theorem
\begin{equation}
\label{eq:bayes}
    \pi(\bm{x}|\bm{D}) = \frac{\mathcal{L}(\bm{D}|\bm{x}) \pi(\bm{x})}{\pi(\bm{D})}
\end{equation}
where $\bm{D}$ is the measured data, and $\pi(\bm{D})=\int \mathcal{L}(\bm{D}|\bm{x}) \pi(\bm{x})d\bm{x} = \mathcal{Z}$. The previous equation includes four major terms that should be defined:
\begin{itemize}
\item $\pi(\bm{x})$: The prior distribution of the input parameters $\bm{x}$, which represents our base knowledge of the tensile cutoff threshold, the density of mercury, and the mercury speed of sound.
\item $\mathcal{L}(\bm{D}|\bm{x})$: The likelihood function, which expresses the likelihood of matching the data at hand $\bm{D}$ given a certain value of the model parameters $\bm{x}$. 
\item $\pi(\bm{x}|\bm{D})$: The updated posterior distribution of the input parameters $\bm{x}$ after combining the new measured data $\bm{D}$.
\item $\pi(\bm{D})$: The marginal likelihood is nothing but a constant ($\mathcal{Z}$), acting as a normalizing factor to ensure that the posterior distribution $\pi(\bm{x}|\bm{D})$ integrates to 1. Therefore, Eq.\eqref{eq:bayes} can be simplified to
\begin{equation}
\label{eq:bayes2}
    \pi(\bm{x}|\bm{D}) = \frac{1}{\mathcal{Z}}\mathcal{L}(\bm{D}|\bm{x}) \pi(\bm{x})
\end{equation}
\end{itemize}

Any mathematical or computer model is a reduced representation of reality due to our incomplete understanding of the underlying physics. Therefore, any model is expected to have an uncertainty that justifies its discrepancy from the reality. Therefore, we define $\epsilon$ as the model discrepancy and write the general relation between the model ($\mathcal{F}$) and the observed data ($\bm{y}$) as 
\begin{equation}
\label{eq:model_eq}
    \bm{y} = \mathcal{F}(\bm{x}) + \epsilon
\end{equation}
where the discrepancy term between the experimental observations and the model predictions can be assumed to follow a Gaussian distribution $\epsilon \sim \mathcal{N}(\epsilon|\bm{0},\bm{\Sigma})$. In practice, the discrepancy term represents the effects of the measurement error (on $\bm{y}$) and the computer/surrogate model inaccuracy. In practical situations like the problem we tackle in this work, it is difficult to know the residual covariance matrix of the discrepancy term ($\bm{\Sigma}$). Alternatively, we can parameterize the covariance matrix by including its unknown parameters as part of the UQ process. Consider a diagonal matrix $\bm{\Sigma}=\sigma^2\bm{I}_{n_y}$ with unknown residual variance $\sigma^2$. The residual variance can be defined as 
\begin{equation}
    \sigma^2 = Var[\epsilon_k], \quad k= 1, ..., n_y.
\end{equation}

We can notice that with this approach, the discrepancy parameter $\epsilon$ reduces to a single scalar representing the variance in the strain data for all 1000 outputs; keeping the inverse UQ problem low-dimensional. The next step is to combine the new added discrepancy parameter to the three original model inputs ($\bm{x}_{\mathcal{F}}$), so the parameter list becomes $\bm{x} = \{\bm{x}_{\mathcal{F}}, \bm{x}_\epsilon \}= \{x_1, x_2, x_3, \sigma^2\}$. Similarly, the new prior distribution for all parameters can be updated as

\begin{equation}
\label{eq:prior}
    \pi(\bm{x}) = \pi(\bm{x}_{\mathcal{F}})\pi(\sigma^2)
\end{equation}

The last component before writing the posterior distribution is the likelihood function, which according to many sources, has a very popular form
\begin{equation}
\label{eq:like}
    \mathcal{L}(\bm{D}|\bm{x}) = \mathcal{L}(\bm{x}_{\mathcal{F}}, \sigma^2;\bm{D}) = \prod_{i=1}^{N_D} \frac{1}{\sqrt{(2\pi\sigma^2)^{n_y}}} exp\Bigg(-\frac{1}{2\sigma^2}[y_i-\mathcal{F}(\bm{x}_{\mathcal{F}})]^T[y_i-\mathcal{F}(\bm{x}_{\mathcal{F}})]\Bigg)
\end{equation}
where $N_D$ is the number of experimental data points $\bm{D}$. Finally, we can write the form of the posterior distribution by plugging back the prior of Eq.\eqref{eq:prior} and the likelihood of Eq.\eqref{eq:like} in the Bayes' formula of Eq.\eqref{eq:bayes2}, yielding
\begin{equation}
\label{eq:post}
    \pi(\bm{x}|\bm{D}) = \pi(\bm{x}_{\mathcal{F}}, \sigma^2|\bm{D}) = \frac{1}{\mathcal{Z}} \pi(\bm{x}_{\mathcal{F}})\pi(\sigma^2) \prod_{i=1}^{N_D} \frac{1}{\sqrt{(2\pi\sigma^2)^{n_y}}} exp\Bigg(-\frac{1}{2\sigma^2}[y_i-\mathcal{F}(\bm{x}_{\mathcal{F}})]^T[y_i-\mathcal{F}(\bm{x}_{\mathcal{F}})]\Bigg).
\end{equation}

Unfortunately, the posterior distribution in Eq.\eqref{eq:post} does not have a closed-form solution, mainly due to the integration that needs to be solved to obtain the marginal likelihood ($\mathcal{Z}$). The most common option to sample from the posterior distribution and hence solving the inverse problem can be done using the Markov Chain Monte Carlo (MCMC) described next. 

\subsection{Markov Chain Monte Carlo (MCMC)}
\label{sec:mcmc}

The idea of MCMC is to construct a Markov chain over the prior distribution to be able to construct an invariant distribution close to the target posterior distribution. The key component of MCMC is the transition probability $\mathcal{K}(\bm{x}^{t+1}|\bm{x}^t)$, which defines the probability to move from the current step at iteration $t$ to the next step at $t+1$ such that this condition is satisfied
\begin{equation}
\label{eq:mcmc}
    \pi(\bm{x}^{t+1}|\bm{D}) \mathcal{K}(\bm{x}^{t}|\bm{x}^{t+1})=\pi(\bm{x}^{t}|\bm{D}) \mathcal{K}(\bm{x}^{t+1}|\bm{x}^t).
\end{equation}

This condition ensures reversibility of the chain, or in other words, the probability to move from $\bm{x}^{t}$ to $\bm{x}^{t+1}$ is equal to the probability to move from $\bm{x}^{t+1}$ to $\bm{x}^{t}$. By integrating Eq.\eqref{eq:mcmc} over the support of the parameters $R_{\bm{x}}$, we can show that the resulting distribution equals to the posterior (i.e. $\int \mathcal{K}(\bm{x}^{t}|\bm{x}^{t+1})d\bm{x}^t = 1$)
\begin{equation}
    \pi(\bm{x}^{t+1}|\bm{D}) = \int_{R_{\bm{x}}} \pi(\bm{x}^{t}|\bm{D}) \mathcal{K}(\bm{x}^{t+1}|\bm{x}^t)d\bm{x}^t.
\end{equation}

Afterward, the analyst can postprocess the MCMC samples to determine the posterior distribution moments (mean, standard deviation, etc.) or fitting the posterior to the closest distribution shape (normal, lognormal, gamma, etc.).

Among the most common MCMC algorithms are the Metropolis–Hastings algorithm, Adaptive Metropolis algorithm, and Hamiltonian Monte Carlo algorithm. However, according to \cite{wagner2019uqlab, goodman2010ensemble}, the affine invariant ensemble algorithm (AIES) \cite{goodman2010ensemble} is more robust to sample from posterior distributions that exhibit strong correlation between its parameters. Therefore, AIES is capable to sample from both types of distributions with or without correlation without explicitly requiring the affine transformation of the target distribution, as required by the prescribed classical MCMC algorithms. Therefore, AIES is used in this work to sample from the posterior distribution of Eq.\eqref{eq:post}. However, it is worth mentioning that AIES is slower and hence more computationally intensive than other MCMC algorithms. 

MCMC algorithms typically require hundreds of thousands of model evaluations, $\mathcal{F}(\bm{x}_{\mathcal{F}})$, before converging to the posterior distribution, see Eq.\eqref{eq:post}. This is why the PCE surrogate model was introduced in section \ref{sec:pce}, as using the real Sierra computer simulator is computationally impractical.  

\subsection{Maximum A Posteriori (MAP)}
\label{sec:map}

The sampling option from the posterior offered by MCMC gives each calibrated parameter in $\bm{x}$ a random distribution of possible values rather than a single point. However, the analyst might be interested in only a single set of parameter values that maximizes the posterior distribution to be used in practice. This approach is called Maximum A Posteriori (MAP)\cite{gauvain1994maximum}, which reports the $\bm{x}$ value that maximizes the posterior distribution as 
\begin{equation}
\label{eq:map}
    \bm{x}_{MAP} = \underset {\bm{x}}{\text{argmax } } \pi(\bm{x}|\bm{D}).
\end{equation}

In this work, we report the MAP values for the tensile cutoff threshold, the density of mercury, and the mercury speed of sound as well as their random distributions found by MCMC.

\subsection{Programming and Computational Tools}

In this work, we use MATLAB as the main programming language. We have also used the mature implementation of the UQLAB to build our PCE surrogate models \cite{marelli2015uqlab} and carry out the Bayesian inverse UQ analysis \cite{wagner2019uqlab}. The UQLAB is a well-validated and reliable framework for surrogate modeling and Bayesian inversion \cite{marelli2014uqlab}.

The Sierra/SolidMechanics 4.56 with EOS material model based on Riemer \cite{riemer2005benchmarking} was implemented in user subroutine VUMAT, which takes about 25 hours on 96 processors (Intel(R) Xeon(R) E-2174G CPU @ 3.80GHz) at a local computer cluster to complete a single simulation. The simulations are majorly conducted using the Argonne Leadership Computing Facility Theta machine to speed up the calculation by leveraging the HPC hardware. The experimental data were collected from actual mercury targets used at the spallation neutron source at the Oak Ridge National Laboratory.

\section{Results and Discussion}
\label{sec:res}

In this section, we first describe the setup and parameter settings used for PCE surrogate modeling and Bayesian inverse UQ analysis. Afterward, the results are presented and discussed simultaneously. 

\subsection{Experimental Setup}

For the PCE surrogate model, 603 simulations have been executed using different combinations of the uncertain parameters. The samples are generated using Latin hypercube sampling with a predefined range for each parameter shown in Table \ref{tab:pce_param} ($x_1$-$x_3$). The wide range for the parameters is assigned to explore the effect of different regions of the three simulation parameters for other purposes outside the scope of this work. The PCE surrogate explores Legendre polynomial orders up to 20, it is trained with 480 samples and validated with 123 samples.

\begin{table}[htbp]
  \centering
  \small
  \caption{Parameters used to construct the PCE surrogate model}
    \begin{tabular}{ll}
    \toprule
    Item  & Value \\
    \midrule
    Number of model inputs ($n_x$) & 3 \\
    Tensile cutoff threshold ($x_1$) & $[0, 1.5\times10^7]$ Pa \\
    Density of mercury ($x_2$) & [1350, 13600] kg/m$^3$ \\
    Mercury speed of sound ($x_3$) & [0, 10000] m/s \\
    Number of model outputs ($n_y$) & 1000 \\
    Maximum polynomial degree ($p_{max}$) & 20 \\
    Polynomial basis ($\phi$) & Legendre \\
    Number of training samples ($N_{train}$) & 480 \\
    Number of validation samples ($N_{val}$) & 123 \\
    \bottomrule
    \end{tabular}%
  \label{tab:pce_param}%
\end{table}%

As this work is probably the first study that digs deeply on the parametric uncertainty of the target simulations, we assume a uniform wide prior for the three simulation parameters as indicated in Table \ref{tab:prior}. The prior range is within the PCE surrogate range in Table \ref{tab:pce_param}, but tighter to represent realistic physics. A very wide prior could have large impact on the Bayesian inference as some of the samples will be lost exploring infeasible posterior regions. For the discrepancy parameter, as assumed by Eq.\eqref{eq:model_eq}, the prior distribution for the residual variance follows a normal distribution. We set a conservative upper bound value of about 1000 $\mu\varepsilon$ (same as $\sigma^2=10^{-6}\varepsilon^2$) for the strain residual variance.  

\begin{table}[htbp]
  \centering
  \small
  \caption{Prior distribution of the model uncertain and discrepancy parameters}
    \begin{tabular}{ll}
    \toprule
    Item  & Value \\
    \midrule
    Tensile cutoff threshold ($x_1$) & $\mathcal{U}[0, 1.5\times10^6]$ Pa \\
    Density of mercury ($x_2$) & $\mathcal{U}[10000, 13600]$ kg/m$^3$ \\
    Mercury speed of sound ($x_3$) & $\mathcal{U}[0, 3000]$ m/s \\
    Discrepancy/Residual variance parameter ($\sigma^2$) & $\mathcal{N}[0, 10^{-6}]$ \\
    \bottomrule
    \end{tabular}%
  \label{tab:prior}%
\end{table}%

The parameter settings for the Bayesian inverse UQ analysis are given in Table \ref{tab:bayes_param}. All measured data introduced in section \ref{sec:data} are used to quantify the uncertainty of the four uncertain parameters (including the residual variance). A total of 120 parallel chains is used for the AIES sampler. Each chain runs for 1200 steps, making the total number of generated samples from the posterior 144,000 samples. We discard the first 72000 samples (i.e. 50\% of the total) and postprocess the other half. This burn-in period is a common practice in MCMC, since the first samples in the chain tend to be of lower quality before the sampler starts to converge. Lastly, we report the MAP value as the point estimate according to section \ref{sec:map}.

\begin{table}[htbp]
  \centering
  \small
  \caption{Parameter settings used in Bayesian inverse UQ}
    \begin{tabular}{ll}
    \toprule
    Item  & Value \\
    \midrule
    Model ($\mathcal{F}$) & PCE surrogate \\
    Number of calibrated parameters & 4 (see Table \ref{tab:prior}) \\
    Number of data points ($N_D$) & 21000 \\
    Number of MCMC chains & 120 \\
    Number of steps per MCMC chain & 1200 \\
    Burn-in samples & 72000 \\
    Point estimate & MAP \\
    \bottomrule
    \end{tabular}%
  \label{tab:bayes_param}%
\end{table}%
   
\subsection{PCE Surrogate Validation}

After the construction of the PCE surrogate using the parameters in Table \ref{tab:pce_param}, it is validated using the validation dataset, and the surrogate metrics are reported in Table \ref{tab:pce_metrics}. The PCE model has a sufficiently low $\epsilon_{LOO}$. According to the UQLAB developers \cite{marelli2015uqlab}, $\epsilon_{LOO} < 0.01$ is considered accurate enough to carry sensitivity analysis and UQ applications. Similarly, the MAE and RMSE are also small. On average, the PCE model can mispredict the strain response by about 32 $\mu \varepsilon$. In addition, having RMSE a bit larger than MAE could imply a large variance in the simulation data where the outliers are penalized by the mean squared error. Overall, it is interesting to observe that PCE maintains a good performance even when the output space is very large ($n_y=1000$). It is worth mentioning here that the PCE model is not supposed to be very accurate since we are not using the surrogate for point-wise predictions, but rather as a tool to accelerate Bayesian analysis. Therefore, capturing the trend of the data could be sufficient for the surrogate as we will be verifying the quality of the Bayesian results by the real computer code at the end. Overall, these metrics boost the confidence in the PCE surrogate model to replace the expensive Sierra computer model when performing Bayesian analysis in the next subsection.

\begin{table}[htbp]
  \centering
  \small
  \caption{PCE surrogate validation metrics (based on the validation set $N_{val}$)}
    \begin{tabular}{ll}
    \toprule
    Item  & Value \\
    \midrule
    Mean Absolute error (MAE) & 32.0 $\mu \varepsilon$ \\
    Root Mean Squared Error (RMSE) & 58.0 $\mu \varepsilon$ \\
    Leave-one-out cross-validation ($\epsilon_{LOO}$) & 3.84E-03 \\
    \bottomrule
    \end{tabular}%
  \label{tab:pce_metrics}%
\end{table}%

\subsection{Inverse UQ Results}

\begin{figure}[!h]
    \centering
    \includegraphics[width=0.8\textwidth]{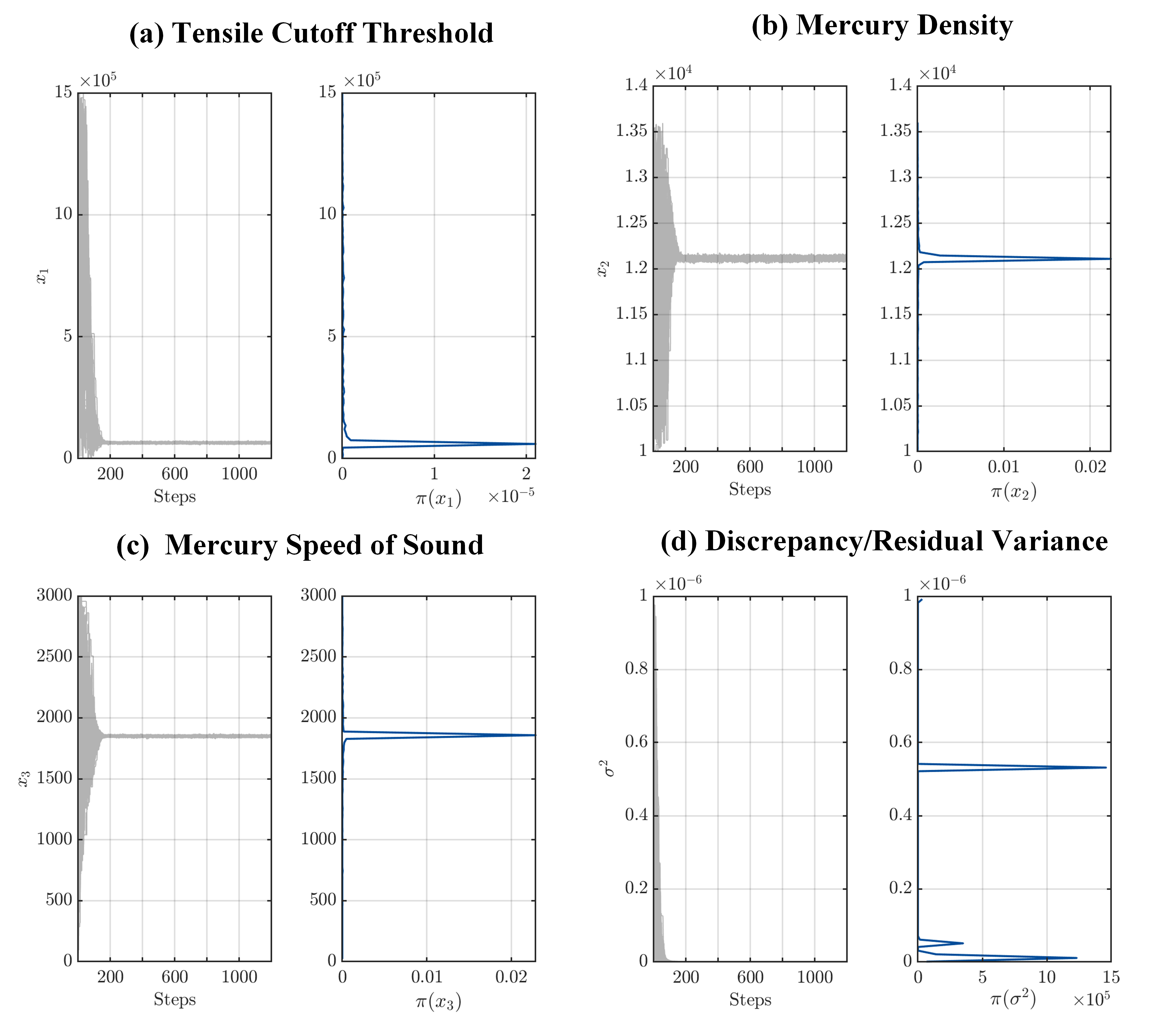}
    \caption{MCMC trace plots of the AIES sampler for (a) tensile cutoff threshold ($x_1$), (b) mercury density ($x_2$), (c) mercury speed of sound ($x_3$), (d) discrepancy/residual variance ($\sigma^2$)}
    \label{fig:trace}
\end{figure}

In this section, we present the MCMC results of sampling from the posterior distribution of Eq.\eqref{eq:post} using the settings provided in Tables \ref{tab:prior}-\ref{tab:bayes_param}. The trace plots of the AIES sampler for the four considered parameters are illustrated in Figure \ref{fig:trace}. For each parameter, the left portion of the subplot shows the trace of the MCMC samples, while the right portion shows the probability density function resulted from these samples. We can notice that the sampler starts by sampling the full range of the parameter space for the first 200 steps to explore the wide prior distribution. Afterward, the sampler starts to converge by identifying the posterior regions that improve fitness to the experimental data. We can observe the sharp peaks in the density function in the right portions of the subplots, which start to develop after 200 steps. The residual variance parameter ($\sigma^2$) features two major peaks at $5.5\times10^{-7}$ and $10^{-9}$, and a small minor peak. The first peak occurred during the early sampling process when the sampler is not stable, as will be seen later in Figure \ref{fig:var}. Therefore the peak at $10^{-9}$ is the most probable. The first impression of the sampling process shows a large reduction in the uncertainty of the three parameters compared to their prior ranges, as we can infer from the tight posterior range of these three parameters. 

After postprocessing the posterior samples by rejecting the first 50\% burn-in samples (see Table \ref{tab:bayes_param}), we provide summary statistics of the calibrated parameters in Table \ref{tab:stat}. The statistics include mean value, standard deviation, the 95\% confidence interval (CI) including lower and upper bounds, and finally the MAP estimate of Eq.\eqref{eq:map}. The inverse UQ results again demonstrate a significant reduction in parametric uncertainty in the three model parameters ($x_1$-$x_3$) as can be inferred from the small standard deviation and the tight confidence intervals in Table \ref{tab:stat}. 

\begin{table}[htbp]
  \centering
  \small
  \caption{Summary statistics for the posterior calibrated parameters and their MAP estimate}
    \begin{tabular}{llllll}
    \toprule
    Parameter & Mean  & Standard Deviation & 95\% CI (Lower) & 95\% CI (Upper) & $\bm{x}_{MAP}$ \\
    \midrule
    $x_1$ & 6.49E+04 & 2.39E+03 & 6.10E+04 & 6.89E+04 & 6.45E+04 \\
    $x_2$ & 12111.8 & 14.9  & 12087.5 & 12136.5 & 12112.1 \\
    $x_3$ & 1849.7 & 5.3   & 1840.9 & 1858.4 & 1850.4 \\
    $\sigma^2$ & 1.66E-09 & 1.56E-11 & 1.64E-09 & 1.69E-09 & 1.66E-09 \\
    \bottomrule
    \end{tabular}%
  \label{tab:stat}%
\end{table}%

\begin{figure}[!h]
    \centering
    \includegraphics[width=0.90\textwidth]{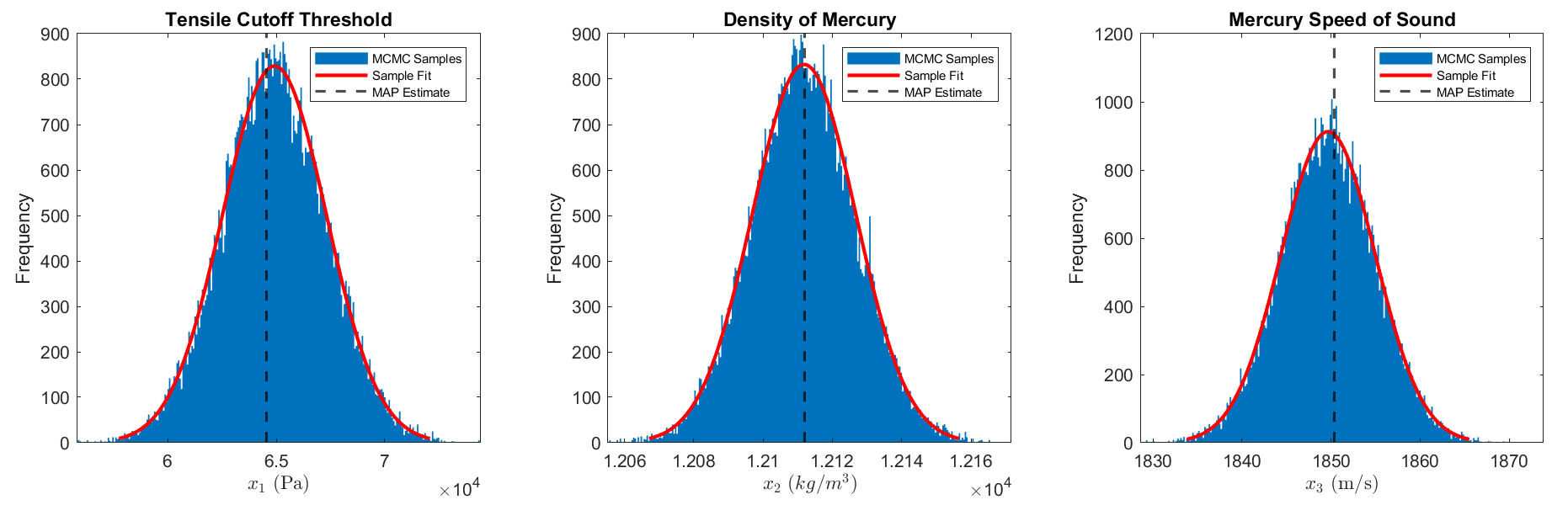}
    \caption{Histogram of posterior samples, fitted marginal distribution, and MAP estimate for (a) tensile cutoff threshold ($x_1$), (b) mercury density ($x_2$), (c) mercury speed of sound ($x_3$)}
    \label{fig:hist}
\end{figure}

Similarly, a plot of the marginal (a.k.a univariate) posterior distribution of the three model parameters can be found in Figure \ref{fig:hist} (residual variance is excluded since it is not a model parameter, but it is shown later in Figure \ref{fig:var}). The posteriors of the three model parameters fit a normal distribution perfectly. In addition, the MAP estimate ($x_{MAP}$) falls in the distribution range (in this case, the normal distribution mode), implying a high accuracy of the MCMC sampling in capturing the posterior marginal.    

As described in section \ref{sec:mcmc}, the AIES sampler is advantageous in sampling the posteriors with strong correlation between their parameters. In Table \ref{tab:corr}, the correlation matrix between the four parameters is presented. First, the correlation is negligible between the discrepancy parameter ($\sigma^2$) and the three model parameters, which should be expected, as the model discrepancy should be independent from the parameter values. Next, the correlation between the model parameters ($x_1-x_3$) is negative and not very strong in general ($ > -0.5$). The largest negative correlation is between tensile cutoff threshold and speed of sound, showing that for a larger speed of sound in mercury, the tensile cutoff is expected to be smaller. When a larger speed of sound in mercury is used, the finite element model tends to drop the density and tensile cutoff threshold values to compensate the mismatch between simulation strain and experimental strain, and vice versa.   

\begin{table}[htbp]
  \centering
  \small
  \caption{Correlation matrix between the four calibrated parameters}
    \begin{tabular}{lllll}
    \toprule
    Correlation & $x_1$ & $x_2$ & $x_3$ & $\sigma^2$ \\
    \midrule
    $x_1$ & 1.00  & -0.18 & -0.43 & 0.00 \\
    $x_2$ & -0.18 & 1.00  & -0.33 & 0.02 \\
    $x_3$ & -0.43 & -0.33 & 1.00  & -0.01 \\
    $\sigma^2$ & 0.00  & 0.02  & -0.01 & 1.00 \\
    \bottomrule
    \end{tabular}%
  \label{tab:corr}%
\end{table}%

\begin{figure}[!h]
    \centering
    \includegraphics[width=0.42\textwidth]{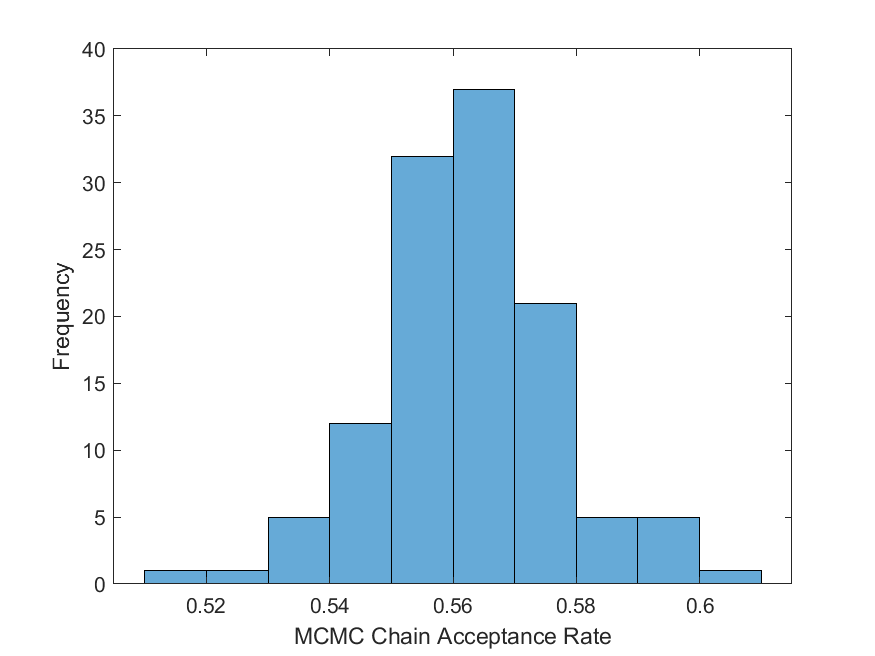}
    \caption{Acceptance rate distribution of the independent MCMC chains}
    \label{fig:acc}
\end{figure}

Lastly, we highlight the acceptance rate of the 120 MCMC chains, which have an average of 0.56 ($56\%$ of the samples were accepted). Although this value is larger than the average acceptance rate for typical MCMC chains (0.25-0.35), we can explain this trend. According to Figure \ref{fig:trace}, a solid convergence of the sampler can be observed after 200 steps. This means that the AIES sampler was likely exploring a tight region of the posterior to search for improvement, which increases the acceptance rate. Furthermore, the histogram in Figure \ref{fig:acc} shows the acceptance rate distribution of the 120 ``independent'' MCMC chains. The distribution shows that all chains have comparable acceptance rates, implying a consistent search trend across all chains. According to \cite{wagner2019uqlab}, it is difficult to assess the quality of MCMC chains solely based on the acceptance rate, but can still be an indicator if the acceptance rate is in a healthy range. Acceptance rate close to 1 one indicates that the MCMC chains did not explore the posterior distribution sufficiently. Acceptance rate close to 0 indicates that the sampled points are in low probability regions of the posterior, so they were rejected. In this work, the acceptance rate falls in the middle, balancing exploration and exploitation of the posterior distribution. 

\subsection{Inverse UQ Validation and Discussion}

\begin{figure}[!h]
    \centering
    \includegraphics[width=0.57\textwidth]{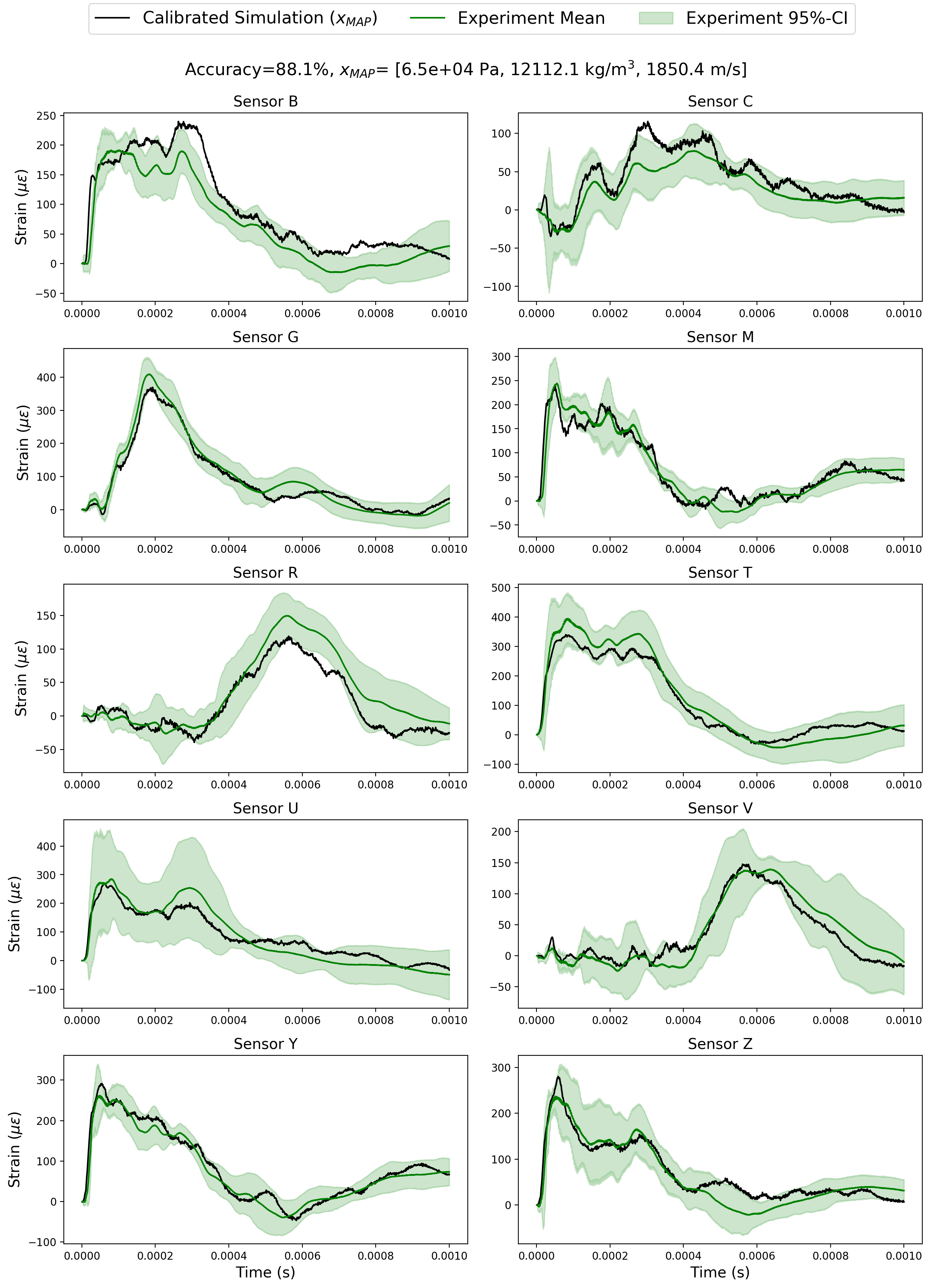}
    \caption{Calibrated simulation against the sensor experimental data using the MAP estimated parameters (see $\bm{x}_{MAP}$ in Table \ref{tab:stat}).}
    \label{fig:calib}
\end{figure}

The inverse UQ findings so far indicate a posterior distribution for each parameter where the analyst can use the mean, median, or mode (e.g. MAP) to continue the analysis based on Table \ref{tab:stat}. Here we continue with $x_1=6.5\times10^4$ Pa for the tensile cutoff threshold, $x_2 = 12112.1$ kg/m$^3$ for the mercury density, and $x_3 = 1850.4$ m/s for the mercury speed of sound. These results are based on the MAP estimate in Table \ref{tab:stat}. To validate our results, we re-run Sierra simulation using these optimal parameter values and compare simulation results to the experimental mean and confidence interval. This will ensure that our quantified parameters can yield a good agreement with the experimental data. Figure \ref{fig:calib} shows a very good agreement between the calibrated simulation and the experimental data for most sensors such as sensors G, M, T, U, V, Y, and Z. As our goal is to ensure the simulation results fall within the experimental confidence interval, we define a quantitative metric called accuracy
\begin{equation}
    Accuracy = \frac{\text{Number of Simulation Points within the Experimental 95\%-CI}}{\text{Total Number of Simulation Points}}\%.
\end{equation}

Table \ref{tab:accuracy} shows the simulation accuracy values for all 10 sensors after using the new simulation parameters. Other than sensor B, the simulation results show outstanding accuracy of 80\% or higher for the remaining sensors with 5 sensors showing 90\% accuracy or higher; validating our inverse UQ findings. The low agreement for sensor B might be due to high strain gradients in the area and potential error in the sensor’s location relative to the simulation point used, as sensor B receives the signal in a diagonal (45$\degree$) form (see Figure \ref{fig:bayes_sensors}). 

\begin{table}[htbp]
  \centering
  \small
  \caption{Simulation accuracy using the MAP parameters (see $\bm{x}_{MAP}$ in Table \ref{tab:stat}) compared to reference parameters by Riemer \cite{riemer2005benchmarking}}
    \begin{tabular}{lll}
    \toprule
    Sensor & Bayesian (\%) & Riemer \cite{riemer2005benchmarking} (\%) \\
    \midrule
    B     & 60    & 70 \\
    C     & 91    & 63 \\
    G     & 89    & 85 \\
    M     & 86    & 75 \\
    R     & 83    & 85 \\
    T     & 100   & 100 \\
    U     & 99    & 100 \\
    V     & 92    & 91 \\
    Y     & 97    & 70 \\
    Z     & 84    & 83 \\
    \textbf{Average} & \textbf{88} & \textbf{82} \\
    \bottomrule
    \end{tabular}%
  \label{tab:accuracy}%
\end{table}%

In addition, Table \ref{tab:accuracy} compares the Bayesian-based simulation accuracy to a reference/baseline parameters reported by Riemer \cite{riemer2005benchmarking}. Our results improved the existing literature knowledge about these physical model parameters with a 6\% increase in average for all sensors' accuracy. Some sensors experienced an improvement of more than 25\% such as sensors C and Y. However, Bayesian inverse UQ tends to sacrifice some sensors to yield a significant improvement in the others, such as sensors R and B. 

As reported by \cite{riemer2005benchmarking}, typical values for these simulation model parameters are: $1.5\times10^5$ Pa for tensile cutoff threshold, $13500$ kg/m$^3$ for mercury density, and $1456$ m/s for mercury speed of sound. In this work, we observed a decrease in mercury density to reflect the heating of the mercury during the experiment, see Figure \ref{fig:bayes_target}(b). Also, it can be observed that mercury speed of sound is a bit higher than the reference value in \cite{riemer2005benchmarking}, and the tensile cutoff is reduced to $6.5\times10^4$ Pa compared to $1.5\times10^5$ Pa in \cite{riemer2005benchmarking}. This is a reflection of the negative correlation we observed before in Table \ref{tab:corr} between $x_1$ and $x_3$. In summary, this new combination of parameters seems to improve the simulation accuracy against the new observed experimental data with an accuracy average of 88\% .  

We noticed in this work that the best mercury density and sound speed that fit the simulation and experiment are a bit off of the nominal mercury physical properties \cite{riemer2005benchmarking}. The reasons for having this offset are coming from three major sources. First, the limitations of the Sierra model or what is known as the ``model-form uncertainty'' that would result from numerical methods and physical approximations. Second reason is the biases and errors in the strain measurements. Third reason is the mercury cavitation damage that also contributes to density reduction. Consequently, the EOS materials model parameters try to compensate for these effects to improve fitness to the data. Therefore, we should emphasize here that those settings should be only used for the problem, computer model, and strain data used in this study, and cannot be generalized for other applications involving liquid mercury. This overfitting nature of the results is indeed the case for most of the Bayesian calibration studies \cite{smith2013uncertainty, kennedy2001bayesian}.  

\begin{figure}[!h]
    \centering
    \includegraphics[width=0.42\textwidth]{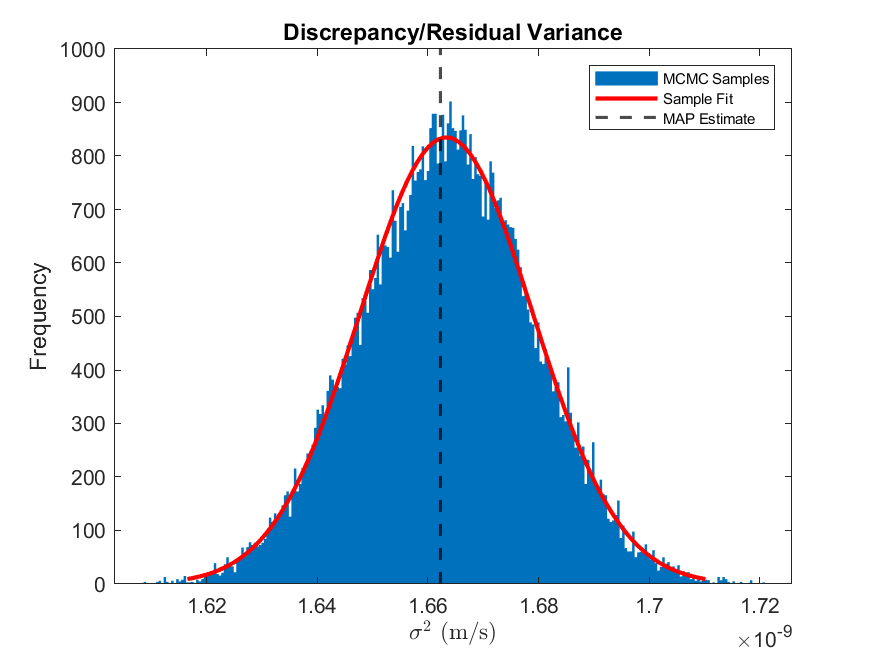}
    \caption{Histogram of posterior samples, fitted marginal distribution, and MAP estimate for the discrepancy/residual variance ($\sigma^2$)}
    \label{fig:var}
\end{figure}

Another interesting finding of this work is the quantification of the experimental uncertainty through the discrepancy parameter or the residual variance. In Figure \ref{fig:var}, we show the histogram of $\sigma^2$ based on the posterior samples. The range and MAP estimate of $\sigma^2$ show an order of $10^{-9}$ for the strain variance. By converting this value to micro-strain to be consistent with other discussions, we find $\sigma = \sqrt{10^{-9}}*10^6 \approx 32 \mu\varepsilon$. Based on this UQ analysis, the value of the upper bound of the measured uncertainty is about $32 \mu\varepsilon$ that yields the best agreement between the experiment and simulation. Nevertheless, this is still an average uncertainty value over all time steps, so the residual variance could vary as a function of the experiment time (e.g. beginning of the pulse, end of the pulse). In addition, this discrepancy includes the error from the surrogate model, and this is why ``upper bound'' is mentioned here for the measured uncertainty. Future efforts may focus on decomposing the total discrepancy to find the surrogate model and the experimental error contributions. Also, assigning a residual variance ($\sigma^2$) for each time step (output) in the experiment is another topic of interest, while keeping in mind that this would make the inverse UQ problem much more complex.

The strain/stress results from finite element simulation are utilized further in fatigue analysis to estimate component’s lifetime and integrity. So, the strain response collected from simulation, particularly the strain peaks and the strain range (max strain - min strain) of the profile are critical in estimating steel vessel’s fatigue cycles, as demonstrated by recent efforts \cite{mach2021fatigue, johns2021design}. At most sensors except R, T and G, the predicted peaks of sensor strains shown in Figure \ref{fig:calib} are close to or higher than the experimental mean values in the short 0.001 second history. Particularly for the sensor B, it has a large over predicted strain peak near the time of 0.3 millisecond, and has the least accuracy (60\%) in predicting full strain profile within 1 millisecond by using the new calibrated parameters set. However these over predicted strain peaks will lead to shorter prediction of target vessel’s lifetime, which tends to be more conservative in target design.

In mercury EOS model, the volumetric stiffness is defined by the bulk modulus, i.e., E\textsubscript{bulk} = density * (sound speed)$^2$. The tensile cutoff threshold defines the point where the tensile pressure is going to be over the threshold value, then it stays at the threshold value while the tensile strain keeps increasing \cite{riemer2005benchmarking}. The new calibrated parameters set results in a higher bulk modulus of (4.144E+10 kg/ms$^2$) than the typical value (2.862E+10 kg/ms$^2$) used in \cite{riemer2005benchmarking}), but a lower tensile cutoff threshold (6.49E+04 vs. 1.5E+05 Pa). This indicates for the new parameters set, the EOS mercury behavior is stiffer to external pressure, but goes into cavitation stage (after tensile cutoff threshold stage) easier than original parameters set in \cite{riemer2005benchmarking}. The new calibrated parameters set is optimized by balancing the accuracy of ten different strain sensors, which may reflect the hidden complex physics in the proton pulsed mercury. They appear to be the best for the current EOS model of \cite{riemer2005benchmarking}, but more efforts are still needed to understand the overall dynamics for building more accurate mercury material model. With more accurate strain response predicted, the component fatigue analysis can utilize the comprehensive strain history data to evaluate target vessel’s lifetime closer to its real limit, saving tremendous target cost and improving design of future targets as well. 

Even so, significant challenges remain. The SNS is undergoing planned upgrade to 2 MW of beam power in the next few years \cite{galambos2020final}, pushing the targets well beyond their currently demonstrated capability. Given that the fabrication time for a target is on the order of two years and the cost is around two million per target vessel, a robust understanding of the physics is desired as well as predictive capabilities on target lifetimes and design modifications.  Unfortunately, diagnostics that can be used to validate existing models are limited to the strain sensor measurements taken over the first few days of target operation for the front sensors, before they fail due to radiation damage, and post-mortem analysis of ``coupons'' removed from the target vessel, which do not directly tie to model output.  In addition, the simulations for the target require intensive computational resources and are not practical for searching large parameter spaces. Therefore, our computational algorithms featuring fast surrogate models and Bayesian inverse UQ to tie the experimental data to the physical parameters are a natural approach to expedite the development of a highly fidelity model that can be used to advance target designs in the future.

\section{Conclusions}
\label{sec:conc}

In this work, we performed an inverse UQ analysis using the Bayesian framework to infer the posterior distribution of physical model parameters of the mercury spallation target. We used Sierra high-fidelity simulations along with measured strain data from 10 sensors to calibrate the tensile cutoff threshold, the density of mercury, and the mercury speed of sound of the Riemer EOS model \cite{riemer2005benchmarking}. To alleviate the computational cost, we replaced the Sierra computer code with a reliable polynomial chaos expansions surrogate model. The inverse UQ analysis demonstrated a maximum-a-posteriori estimate of $6.5\times10^4$ Pa for tensile cutoff threshold, $12112.1$ kg/m$^3$ for mercury density, and $1850.4$ m/s for mercury speed of sound, and all three parameters were closely fitted by a normal distribution. These new parametric values result in an improvement in Sierra simulations with an average of 88\% accuracy compared to experimental data, and a 6\% average increase in accuracy compared to the reference parameters by \cite{riemer2005benchmarking}, with some sensors experiencing more than 25\% increase in accuracy. Explanations of the new calibrated values that are different from nominal parameters include the limitations of the computer model (model-form uncertainty) due to numerical methods and physical approximations, the biases and errors in the experimental data, and the mercury cavitation damage that also contributes to the change in mercury behaviour. With more accurate simulations of the strain response, the component fatigue analysis can utilize the comprehensive strain history data to evaluate target vessel’s lifetime, saving tremendous target costs. Our future research will focus on scaling this methodology to more advanced mercury target simulations that exhibit more complex physics and data (i.e. helium gas injection in the target during the pulse) as well as more uncertain parameters (i.e. 8-10 parameters).

\section*{Acknowledgment}

The authors are grateful for support from the Neutron Sciences Directorate at ORNL in the investigation of this work. This work was supported by the DOE Office of Science under grant DE-SC0009915 (Office of Basic Energy Sciences, Scientific User Facilities program). This research also used resources of the Argonne Leadership Computing Facility, which is a DOE Office of Science User Facility supported under Contract DE-AC02-06CH11357.

Notice: This manuscript has been authored by UT-Battelle, LLC, under contract DE-AC05-00OR22725 with the US Department of Energy (DOE). The US government retains and the publisher, by accepting the article for publication, acknowledges that the US government retains a nonexclusive, paid-up, irrevocable, worldwide license to publish or reproduce the published form of this manuscript, or allow others to do so, for US government purposes. DOE will provide public access to these results of federally sponsored research in accordance with the DOE Public Access Plan (http://energy.gov/downloads/doe-public-access-plan).

\section*{CRediT Author Statement}

\noindent \textbf{Majdi I. Radaideh}: Conceptualization, Methodology, Software, Validation, Investigation, Data curation, Visualisation, Formal analysis, Writing - Original Draft. \\
\textbf{Lianshan Lin}: Conceptualization, Methodology, Software, Validation, Writing – Review and Edit. \\
\textbf{Hao Jiang}: Conceptualization, Data curation, Writing – Review and Edit. \\
\textbf{Sarah Cousineau}: Conceptualization, Funding acquisition, Resources, Writing – Review and Edit.


\bibliographystyle{elsarticle-num}
\setlength{\bibsep}{0pt plus 0.3ex}
{
\footnotesize \bibliography{references}}

\end{document}